\documentclass[12pt]{article}
\usepackage{graphicx}

\begin{document}
\title{ANCOVA: A GLOBAL TEST BASED ON A ROBUST MEASURE OF LOCATION OR QUANTILES  WHEN THERE IS CURVATURE}
\author{Rand R. Wilcox \\
Dept of Psychology \\
University of Southern California
}
\maketitle
\pagebreak
\begin{center} 
ABSTRACT
\end{center}

For two independent groups, let $M_j(x)$ be some conditional measure of location for the $j$th group associated with some random variable $Y$, given that some covariate $X=x$. When $M_j(x)$
is a robust measure of location, or even some conditional quantile  of $Y$, given $X$, methods have been proposed and  studied that are aimed at testing $H_0$: $M_1(x)=M_2(x)$
that deal with curvature in a flexible manner. In addition, methods have been studied where the goal is to control the probability of one or more Type I errors when
testing $H_0$ for each $x \in \{x_1, \ldots, x_p\}$. This paper  suggests a method for testing the global hypothesis  $H_0$: $M_1(x)=M_2(x)$ for $\forall x  \in \{x_1, \ldots, x_p\}$
when using a robust or quantile location estimator.
An obvious advantage of testing $p$ hypotheses, rather than the global hypothesis, is that it can provide information about where regression lines differ and by how much.
But the paper summarizes three general reasons to suspect that testing the global hypothesis can have more power.
 Data
from the Well Elderly 2 study illustrate that testing the global hypothesis can make a practical difference.


\section{Introduction}

For two independent groups,  consider the  situation  where for the $j$th group ($j=1$, 2)
$Y_j$ is  some outcome variable of interest  and $X_j$ is some covariate.  The classic ANCOVA method assumes that 
\begin{equation}
Y_j = \beta_{0j} + \beta_1X_j + \epsilon, \label{usual}
\end{equation}
where $\beta_{0j}$ and $\beta_1$ are unknown parameters and $\epsilon$ is a random variable having a normal distribution with mean zero and unknown variance $\sigma^2$.
So the regression lines are assumed to be parallel and the goal is to compare the intercepts  based in part on a least squares estimate of the regression lines. 
It is well known, however, that there are serious concerns with this approach. First, there is a vast literature establishing that methods based on means, including 
least squares regression, are not robust (e.g.,  Staudte and Sheather, 1990; 
 Marrona et al., 2006; 
Heritier et al., 2007; Hampel et al., 1986;
Huber and Ronchetti, 2009;  Wilcox, 2012a, 2012b).  A general concern is that violations of underlying assumptions  can result in relatively poor power and poor control over the
Type I error probability. Moreover, even a single outlier can yield  a poor fit to the bulk of the points when using least squares regression. 

As is evident, one way of dealing with non-normality is to use some rank-based technique.  But rank-based ANCOVA methods are aimed
a testing the hypothesis of identical distributions  (e.g., Lawson, 1983).  So when this method rejects, it is reasonable to conclude that the distributions differ in some manner,
but the details regarding how they differ, and by how much, are unclear. One way of gaining some understanding of how the groups differ, but certainly not the only way, is to
compare the groups using some measure of location. Here the goal is to make inferences
about some robust (conditional) measure of location associated with $Y$.  

Yet another fundamental concern with (\ref{usual}) is that the true regression lines are assumed to be straight. Certainly, in some situations, this is a reasonable approximation.
 When there is curvature, simply meaning that the regression line is not straight, using some obvious parametric regression model might suffice. (For example, include a 
quadratic term.) But this approach can be inadequate, which has led  to a substantial collection of nonparametric methods, often called smoothers, for dealing with curvature in a more
flexible manner
(e.g., H\"{a}rdle, 1990;  Efromovich, 1999;  Eubank , 1999; Fox, 2001; Gy\"{o}rfi, et al., 2002).  


Here, the model given by (\ref{usual})  is replaced with the less restrictive model
 \begin{equation}
Y_j = f_j(X_j) + \epsilon_j, \label{model}
\end{equation}
where  $f_j(X_j)$ is some unknown  function that reflects some conditional measure of location associated with $Y$ given that the covariate value is $X_j$. The random
variable  $\epsilon_j$ has some unknown distribution with variance $\sigma^2_j$. 
So unlike the classic approach where it is assumed that 
\[ f_j(X_j) =  \beta_{0j} + \beta_{1j}X_j, \]
no parametric model for $f_j(X_j)$  is specified and  $\sigma^2_1=\sigma^2_2$ is not assumed.
Let $M_j(x)$ be some (conditional) measure of location associated with $Y_j$ given that $X_j=x$.
Here,  curvature  is addressed using a running interval smoother. Roughly, like all smoothers, the basic strategy is to focus on  the 
$X_j$ values close to $x$ and use the corresponding  $Y_j$ values to  estimate  $M_j(x)$. An appeal of the running interval smoother is that it 
is easily applied when using any robust measure of location. The details are given in the next section of this paper. 

The goal here is  to test  the global hypothesis
\begin{equation}
H_0: M_1(x)=M_2(x), \, \forall x \in \{x_1, \ldots, x_p\}, \label{null_glob}
\end{equation}
where $x_1, \ldots, x_p$ are $p$ values of the covariate chosen empirically  in a manner  aimed at capturing any
 curvature that might exist.  Roughly, these $p$ values are chosen using a component of the so-called running interval smoother, which is described in section 2.
 Put in more substantive terms, the goal is to determine whether  two groups differ 
  (e.g., depressive symptoms among males and females)  taking into account the possibility that the extent they differ might depend in a non-trivial manner 
 on some covariate (such as the cortisol awakening response). 
 
  In the context of ANCOVA, use of the 
running interval smoother is not new.  In particular Wilcox (1997) proposed and studied a method that tests $H_0$: $M_1(x_k)=M_2(x_k)$
for each $k$, $k=1, \ldots, p$. So $p$ hypotheses are tested rather than the global hypothesis corresponding to (\ref{null_glob}). 
The method is based in part on Yuen's (1974) method for comparing trimmed means with the familywise error rate (the probability of one or more Type I errors) controlled
using a strategy that is similar to Dunnett's (1980) T3 technique. 
More recently, a bootstrap variation
 was proposed and studied by Wilcox (2009).  Now the familywise error rate can be controlled using some improvement on the Bonferroni method (e.g., Rom, 1990; Hochberg, 1988).
 The bootstrap method can, in principle, be used with any robust measure of location.

However, a practical concern with testing $p$ individual hypotheses, rather than a global hypothesis, is that power might be relatively low for three general reasons. 
First, each individual hypothesis uses only a subset of the available data. In contrast, the global hypothesis used here is based on all of the data that are used to test the individual hypotheses.
That is, a larger sample size is used suggesting that
it might reject in situations where the none of individual tests is significant. Second, if for example the  familywise error rate is set at .05, then Wilcox's method uses a Type I error probability
less than .05 for the individual tests, which again can reduce power. 
 The third reason has to do with using a confidence region for two or more parameters as opposed to confidence intervals
for each individual parameter of interest. It is known that in various situations, confidence regions can result in a significant difference even when there are 
non-significant results  for the individual parameters.
 (For an illustration, see for example Wilcox, 2012b, p. 690.)
The method proposed here
 for testing (\ref{null_glob}) deals with this issue in a manner that is made clear in section 3. 
  Data from the Well Elderly 2 study (Jackson et al., 2009; Clark et al., 2011) are used to illustrate that the new method can make a practical difference. 

Another goal in this paper  is to include simulation results on comparing (conditional) quartiles. Comparing medians is an obvious way of 
proceeding. But   
in some situations, differences in the tails of two distributions can be more important and informative than comparisons based on a measure of location that is 
centrally located (e.g., Doksum \& Sievers, 1976; Lombard, 2005). This proved to be the case in the Well Elderly 2 study for reasons explained  
 in section 4. 
 
Note that rather than testing (\ref{null_glob}), a seemingly natural goal  is to test the hypothesis that $M_1(x)=M_2(x)$ for all possible values of  $x$, not just those values in the set
$ \{x_1, \ldots, x_p\}$. 
Numerous papers contain results on methods for accomplishing this goal when $M_j(x)$ is taken to be the conditional mean of $Y$ given that $X=x$.
 (For a list of references, see Wilcox, 2012a, p. 610.)
But the mean is not robust and evidently little or nothing is known about how best to proceed when using some robust measure of location. 
Wilcox (2012a, section 11.11.5) describes a robust method based on a running interval smoother, but the choice for the span (the value of $\ell_j$ described in the next section)
is dictated by the sample size given the goal of controlling the Type I error probability. That is, a suboptimal fit to the data might be needed. The method used here avoids this problem.
Here, some consideration was given to an approach where a robust smoother is applied to each group and predicted $Y$ values are computed for all of the observed  $x$ values. 
If the null hypothesis is true, the regression line for 
the differences $M_1(x)-M_2(x)$,  versus $x$,  should have a zero slope and intercept. 
Several bootstrap methods were considered based on this approach, but control over the Type I error probability was very poor,
so no details are provided.  


\section{Description of the Proposed Method}

Following Wilcox (1997), the general strategy is to approximate the regression lines with a
running interval smoother  and then use
the components of the smoother to test some relevant hypothesis.  
A portion of the method requires choosing a location estimator. 
As will be made evident, in principle any robust location estimator could be used, but here attention is focused on only two estimators: 
a 20\% trimmed mean and 
the quantile estimator derived by Harrell and Davis (1982). 

Let  $Z_1, \ldots, Z_n$ be any $n$ observations. The $\gamma$-trimmed mean is
\[\frac{1}{n-2g} \sum_{i=g+1}^{n-g} Z_{(i)},\]
where $Z_{(1)} \le \cdots \le Z_{(n)}$ are the $Z$ values written in ascending order and $g=\lfloor \gamma n \rfloor$
is the greatest integer less than or equal to $\gamma n$, $0 \le \gamma < .5$.  The 20\% trimmed mean corresponds to  $\gamma=.2$. 
One advantage of the 20\% trimmed mean is that its efficiency
compares well to the sample mean under normality (e.g., Rosenberger \& Gasko, 1983). But  as we
move toward a more heavy-tailed distribution, the standard error of the  20\% trimmed mean can be substantially smaller than the
standard error of the mean, which can translate into substantially higher power when outliers tend to occur. 
Another appeal of the 20\% trimmed mean over the mean, when testing hypotheses,  is that both theory and simulations indicate that the 20\% trimmed is better at handling skewed
distributions in terms of controlling the Type I error probability. 
This is not to suggest that the 20\% trimmed mean
dominates all other
robust estimators that might be used.  Clearly this is not the case. The only point is that it is a reasonable measure of location to consider for the situation at hand.

The Harrell and Davis (1982)  estimate of the $q$th quantile uses a 
 weighted average of all the order statistics. 
 Let $U$ be a random variable having a beta distribution with parameters
$a=(n+1)q$ and $b=(n+1)(1-q)$ and let
 \[v_i = P\left(\frac{i-1}{n} \le U \le \frac{i}{n}\right).\]
The estimate of the $q$th quantile, based on $Z_1, \ldots, Z_n$, is
\begin{equation}
\hat{\theta}_{q} = \sum_{i=1}^n v_i Z_{(i)}. \label{hdest}
\end{equation}

In terms of its standard error, Sfakianakis and Verginis (2006) show that in some situations the Harrell--Davis estimator competes well with alternative 
estimators that again use a weighted average of all the order statistics, but there are exceptions.
(Sfakianakis and Verginis derived alternative estimators that have advantages over the Harrell--Davis in some situations. But we found that
when sampling from heavy-tailed distributions, the standard error of their estimators can be substantially larger than the standard error of $\hat{\theta}_{q}$.) Comparisons with other quantile
 estimators
are reported by Parrish (1990),
Sheather and Marron (1990),
as well as Dielman, Lowry and Pfaffenberger (1994). 
 The only certainty is that no single estimator dominates in terms
 of efficiency. For example, the Harrell--Davis estimator has a smaller standard error than the usual sample median when sampling from a normal
 distribution or a distribution that has relatively light tails, but for sufficiently heavy-tailed distributions, the reverse is true (Wilcox, 2012a, p. 87).

To describe the details of the method for testing (\ref{null_glob}),  let $(X_{ij}, Y_{ij})$  ($i=1, \ldots, n_j$;  $j=1$, 2) be a random sample of size $n_j$ from the $j$th group.   For 
a chosen value for $x$, suppose the goal is to estimate $M_j(x)$.
The strategy is simple. Roughly,  for each $j$, compute a measure of location based on the $Y_{ij}$ values 
for which the corresponding $X_{ij}$ values are close to $x$. More formally, for fixed $j$, compute a measure of location based on the $Y_{ij}$ values 
 such
that $i$ is an element of the set
\[P_j(x) = \{i:\,|X_{ij}-x| \le \ell_j \times {\rm MADN}_j\},\]
where $\ell_j$ is a constant chosen by the investigator and often called the span, MADN$_j$=MAD$_j$/.6745,  MAD$_j$ 
(the median absolute deviation) is the median of $|X_{1j}-m_j|, \ldots |X_{n_jj}-m_j|$ and $m_j$ is the usual
sample median based on $X_{1j}, \ldots, X_{n_jj}$. Under normality, MADN$_j$=MAD$_j$/.6745 estimates the population standard deviation, in which case $X_{ij}$ is close to 
$x$ if it is within  $\ell_j$ standard deviations from $x$.
Generally, the choice  $\ell_j=.8$ or $\ell_j=1$  gives
good results, in terms of capturing any curvature,  but  of course exceptions are encountered. 

Let $N_j(x)$ be the cardinality of the set $P_j(x)$ and suppose that $M_j(x)$ is estimated with some measure of location based on the 
$Y_{ij}$ values for which $i \in P_j(x)$.    
The two regression lines are defined
to be  comparable at $x$ if simultaneously $N_1(x) \ge 12$ and $N_2(x) \ge 12$.
The idea is that if the sample sizes used to estimate $M_1(x)$ and
$M_2(x)$ are
sufficiently large, then a reasonably accurate confidence interval for
$M_1(x)-M_2(x)$ can be computed provided a reasonably level robust technique is used. For example, Yuen's (1974)  method might be used with a 
20\% trimmed mean. (It is known that under fairly general conditions methods for comparing means are not level robust with
relatively small sample sizes.  See Wilcox, 2012b, for details.)

For notational convenience, let $\hat{\theta}_{jk}$ be some location estimator  based on the $Y_{ij}$ values for which
 $i \in P_j(x_k)$. Let $\hat{\delta}_k=\hat{\theta}_{1k}-\hat{\theta}_{2k}$ and let $\delta_k$ denote the population analog of $\hat{\delta}_k$ ($k=1, \ldots, p$). Then (\ref{null_glob}) 
 corresponds to 
 \begin{equation}
H_0:  \delta_1=\delta_2= \cdots \delta_p=0. \label{nullalt}
\end{equation}
 The basic strategy for testing (\ref{nullalt}) is to generate bootstrap samples from each group, compute $\hat{\delta}_k$ based on these bootstrap samples, repeat this $B$ times, and
 then measure how deeply the null vector ${\bf 0}$ is nested in the bootstrap cloud of points via Mahalanobis distance.
 Based on these distances, results in Liu and Singh (1997) indicate how to compute a p-value.
 
To elaborate, let $(X^*_{ij}, Y^*_{ij})$  be a bootstrap sample from the $j$th  group, which is obtained by resampling with replacement $n_j$ pairs of points from $(X_{ij}, Y_{ij})$
($i=1, \ldots, n_j$; $j=1$, 2).  Let $\hat{\delta}_k^*$ be the estimate of  $\delta_k$ based on the bootstrap samples from the two groups. Repeat this process $B$ times
yielding  $\hat{\Delta}^*_b =(\hat{\delta}_{1b}^*, \ldots, \hat{\delta}_{pb}^*)$, $b=1, \ldots, B$.
 Let ${\bf S}$ be the covariance matrix based on the $B$ vectors $\hat{\Delta}^*_{1}, \ldots,  \hat{\Delta}_{B}^*$.
 Note that the center of the bootstrap cloud being estimated by these $B$ bootstrap samples is known. It is $\hat{\Delta} = (\hat{\delta}_1, \ldots, \hat{\delta}_p)$, the estimate
 of $\Delta = (\delta_1, \ldots, \delta_p)$ based on the $(X_{ij}, Y_{ij})$ values. 
Let
\[d^2_b = (\hat{\Delta}^*_b - \hat{\Delta})  {\bf S}^{-1}  (\hat{\Delta}^*_b - \hat{\Delta})^{\prime},\]
where for $b=0$,   $\hat{\Delta}^*_0$ is taken to be the null vector  ${\bf 0}$.  Then a (generalized) p-value is 
\begin{equation}
 \frac{1}{B} \sum_{b=1}^B  I(d^2_0 \le d^2_b), \label{pv}
 \end{equation}
where the indicator function $I(d^2_0 \le d^2_b)=1$ if $d^2_0 \le d^2_b$; otherwise $I(d^2_0 \le d^2_b)=0$.

There remains the problem of choosing the $x_k$ values. They might be chosen based on substantive grounds, but of course studying this strategy via simulations is
difficult at best. Here, we follow Wilcox (1997) and choose $p=5$ points in a manner suggested by running interval smoother in terms of capturing any curvature in a flexible manner. 
For notational
convenience, assume that for fixed $j$, the $X_{ij}$ values are
in ascending order. That is, $X_{1j} \le \cdots \le X_{n_{j}j}$.
Suppose
$z_1$ is taken to be the smallest   $X_{i1}$ value for which
the regression  lines are comparable. That is, search the first
group for the smallest $X_{i1}$ such that $N_1(X_{i1}) \ge 12$.
If $N_2(X_{i1}) \ge 12$, in which case the two regression lines
are comparable at $X_{i1}$, set $x_1=X_{i1}$.
If $N_2(x_{i1}) < 12$,
consider the next largest $x_{i1}$ value and continue until
it is simultaneously true that
$N_1(X_{i1}) \ge 12$ and $N_2(X_{i1}) \ge 12$.
Let $i_1$ be the smallest integer such that
$N_1(x_{i_11}) \ge 12$ and $N_2(x_{i_11}) \ge 12$.
Similarly, let $x_5$ be the largest $X_{i1}$ value for which the regression lines are comparable. That is,
$x_5$ is the largest $X_{i1}$ value such that $N_1(x_{i1}) \ge 12$ and
$N_2(x_{i1}) \ge 12$. Let $i_5$ be the corresponding value of $i$.
Let $i_3=(i_1+i_5)/2$, $i_2 = (i_1+i_3)/2$, and $i_4=(i_3+i_5)/2$.
Round $i_2$, $i_3$, and $i_4$ down to the nearest integer and set
$x_2=X_{i_21}$, $x_3=X_{i_31}$, and $x_4=X_{i_41}$. 

When the covariate values are chosen in the manner just described, and $p=5$ separate tests are performed based on some measure of location, this will be called method W henceforth.
Computing a p-value using (\ref{pv}), with the goal of performing a global test, will be called method G.  Unless stated otherwise, both methods G and W will be based on a 20\% trimmed mean.

Note that in essence, we have a 2-by-p ANOVA design. But for the p levels of the second factor,  the groups are not necessarily  independent. The reason is that for any two covariate values,
say $x_k$ and $x_m$, the intersection of the sets $P_j(x_k)$ and $P_j(x_m)$ is not necessarily equal to the empty set. Here, the strategy for dealing with this feature is 
to model it via a bootstrap method. 
Another approach would be divide the data into $p$ independent groups. But  there is uncertainty about how this might be done so as to effectively  capture any curvature. The approach used here mimics 
a basic component used by a wide range of smoothers designed to deal with curvature in a flexible manner. 
 
Of course, the obvious decision rule, when using method G, is to reject the null hypothesis if the p-value is less than or equal to the nominal level. When testing at the $\alpha=.05$ level,
preliminary simulations indicated that this approach performs well, in term of controlling the Type I error probability, when
$p=3$ and the $x_k$ values are taken to be the quartiles corresponding to the $X_{i1}$ values.  But when $p=5$
and the $x_k$ values are chosen as just described, the actual level exceeded .075 when testing at the $\alpha=.05$ level with $n_1=n_2=30$. This problem persisted with
$n_1=n_2=50$. However, for the range of distributions considered (described in section 3), the actual level was found to be relatively stable. This suggests using a
strategy similar to Gosset's (Student's) approach to comparing means: Assume normality, determine an appropriate critical value using a reasonable test statistic, and continue
using this critical value when dealing with non-normal distributions.

Given $n_1$ and $n_2$, 
this strategy is implemented by first by generating, for each $j$, $n_j$ pairs of observations  from a bivariate normal distribution having a correlation 
$\rho=0$. Based on this generated data, determine  $p=5$ values of the covariate in the manner just described and then compute the p-value given by (\ref{pv}).
Denote this p-value by $\hat{p}$. Repeat this process $A$ times yielding  $\hat{p}_1, \ldots, \hat{p}_A$. Then an $\alpha$ level critical p-value, say $\hat{p}_c$, is taken to be
the $\alpha$ quantile of the $\hat{p}_1, \ldots, \hat{p}_A$ values, which here is estimated via the Harrell-Davis estimator. (With A=1000 and when a trimmed mean is used,
this can be done in 14.8 seconds using
an R function, described in the final section of this paper, running on the
first author's MacBook Pro.)  That is, letting $p_o$ denote the p-value based on the observed data, reject (\ref{null_glob}) if $p_o \le \hat{p}_c$.

Note that once $p_c$ has been determined, a  $1-\alpha$ confidence region for the vector  $\Delta=(\delta_1, \ldots \delta_p)$ can be computed. A confidence region consists of the convex hull
containing the $(1-\hat{p}_c)B$  $\hat{\Delta}_b$ vectors that have the smallest $d^2_b$ values.  As previously indicated, 
this confidence region provides a perspective on why the global test considered here
can have more power than method W. Situations are encountered where the null vector is not contained in the confidence region, yet the confidence intervals for each
of the $p$ differences contain zero. 

\section{Simulation Results}


Simulations were used to study the small-sample
properties of the proposed method with $n_1=n_2=30$.  Smaller sample sizes are dubious because this 
makes it particularly difficult to effectively deal with
curvature. Also,  finding five covariate values where the 
groups are comparable can be problematic. That is, $N_j(x)$  might be so small as to make comparisons meaningless.
A few results are reported with $n_1=n_2=100$ and 200 as well.
 
Estimated Type I error probabilities, $\hat{\alpha}$, were based on 4000 replications.
The estimated critical p-value was based on $A=1000$ and $B=500$ bootstrap samples.
Four types of distributions were used:
normal, symmetric and heavy-tailed, asymmetric and light-tailed,
and asymmetric and heavy-tailed.
More precisely, the marginal distributions were taken to be one of four g-and-h distributions
(Hoaglin, 1985) that contain the standard  normal distribution as a special case. (The R function ghdist, in Wilcox, 2012a, was used to generate
observations from a g-and-h distribution.)
If $Z$ has a standard normal distribution, then by definition
\[V = \left\{ \begin{array}{ll}
 \frac{{\rm exp}(gZ)-1}{g} {\rm exp}(hZ^2/2), & \mbox{if $g>0$}\\
  Z{\rm exp}(hZ^2/2), & \mbox{if $g=0$}
   \end{array} \right. \]
has a g-and-h distribution where $g$ and $h$ are parameters that
determine the first four moments. That is, a g-and-h distribution is a transformation of the standard normal random variable that can be used to
generate data having a range of skewness and kurtosis values. 
The four distributions used here were the standard normal ($g=h=0.0$), a
symmetric heavy-tailed distribution ($h=0.2$, $g=0.0$), an asymmetric
distribution with
relatively light tails ($h=0.0$, $g=0.2$), and an asymmetric distribution with
heavy tails ($g=h=0.2$).
Table 1 shows the skewness ($\kappa_1$) and kurtosis
($\kappa_2$)
for each distribution. Additional properties of the g-and-h distribution
are summarized by Hoaglin (1985).

\begin{table}
\caption{Some properties of the g-and-h distribution.}
\centering
\begin{tabular}{ccrr} \hline
g & h &  $\kappa_1$ & $\kappa_2$\\ \hline
0.0 & 0.0  & 0.00 & 3.0\\
0.0 & 0.2 & 0.00 & 21.46\\
0.2 & 0.0  & 0.61 & 3.68\\
0.2 & 0.2 & 2.81 & 155.98\\ \hline
\end{tabular}
\end{table}

The g-and-h distributions with  $h=.2$ were chosen in an attempt to span the range of distributions that might be encountered in practice. 
The idea is that if method G performs well for what some might regard as an unrealistic departure from normality,
 this provides some reassurance that it will perform reasonably when
dealing with data from an actual study.

Three types of associations were considered. The first two deal with situations where $Y_{ij}=\beta X_{ij}+ \epsilon$. The two choices for the slope
were $\beta=0$ and 1.  The third  type of association was
$Y_{ij}= X_{ij}^2+ \epsilon$.  These three situations are labeled S1, S2 and S3, respectively. The estimated Type I errors were very similar for S1 and S2, 
so for brevity the results for  S2 are not reported. The $X_{ij}$ values were generated from a standard normal distribution and $\epsilon$ was generated from one
of the four g-and-h  distributions previously indicated.

The simulation results are reported in Table 2. As can be seen, when testing at the .05 level, the actual level was estimated to be less than or equal to
 .056 among all of the
situations considered. 
Although the seriousness of a Type I error depends on the situation, Bradley (1978) suggests that as a general guide, when testing at the .05 level, the actual level should
be between .025 and .075. Based on this criterion, 
the only concern is that for a very heavy-tailed distribution, the estimated level drops below .025, the lowest estimate being .020. Increasing both sample sizes to 50 corrects this problem.
For example, with $g=h=.2$ and $\gamma=.2$, the estimate for situation S1 increases from .020 to .034. 

\begin{table}
\center
\caption{Estimated  Type I error probabilities when testing at the $\alpha=.05$ level, $n_1=n_2=30$}
\begin{tabular}{cccccc}
$g$ & $h$ & Estimator & S1 & S3\\ \hline
0.0 & 0.0 &  $\gamma=.2$ &   .048 & .048\\   
0.0 & 0.0 & $q=.50$ &   .038 & .044 \\   
0.0 & 0.0 & $q=.75$ &   .049 & .048 \\    

0.0 & 0.2 &  $\gamma=.2$ &   .022 & .026\\  
0.0 & 0.2 & $q=.50$ &   .023 & .028 \\   
0.0 & 0.2 & $q=.75$ &   .029 & .028 \\

0.2 & 0.0 &  $\gamma=.2$ &   .040 & .047\\  
0.2 & 0.0 & $q=.25$ &   .053 & .056\\  
0.2 & 0.0 & $q=.50$ &   .036 & .044 \\ 
0.2 & 0.0 & $q=.75$ &   .046 & .045 \\

0.2 & 0.2 &  $\gamma=.2$ &   .020 & .024\\  
0.2 & 0.2 & $q=.25$ &   .040 & .040\\  
0.2 & 0.2 & $q=.50$ &   .022 & .028 \\   
0.2 & 0.2 & $q=.75$ &   .026 & .025 \\

\hline

\end{tabular}
\end{table}

Notice that the lowest estimates in Table 2 occur for  $\gamma=.2$  when $g=h=.2$. Simulations were run again with $n_1=n_2=100$ as well as $n_1=n_2=200$ as a partial check
on the impact of using larger sample sizes. The estimated Type I error probabilities for these two situations were .036 and .040, respectively. 



As previously explained, there are at least three reasons to expect that  the global test  will have more power than  method  W.
The extent this is true depends on the situation. 
To provide at least some perspective, consider the case where  the covariate has a normal distribution and the error term has 
a g-and-h distribution. First consider $g=h=0$ (normality),  and suppose the first group has $\beta_1=\beta_0=0$, while for the second group $Y=.5+ \epsilon$.
With $n_1=n_2=50$, and testing at the .05 level, the power of method G test was estimated to be .51. 
The probability of rejecting at one or more
design points using  method W was estimated to be  .38.  If instead  $Y=.5X+.5+ \epsilon$ for the second group, the power estimates are now
.75 and .66, respectively. If  $Y=.5X^2+.5+ \epsilon$, the estimates are .89 and .78. For this last situation, if  $(g, h)=(0,.2)$, the estimates are .76 and .70.
For  $(g, h)=(.2,.2)$ the estimates are .75 and .70. So all indications are that W has more power, with the increase in power estimated to be as high as .12 among
the situations considered here. 

As already noted, a well-known argument for using a 20\%  trimmed mean, rather than the mean, is that under normality its efficiency compares very well to to the mean, but as we move toward a 
heavy-tailed distribution,  the standard error of the mean can be substantially larger than the standard error of the 20\% trimmed. 
That is, in terms of power, there is little separating the mean and 20\% under normality, but for heavier tailed distributions, power might be substantially higher using a  20\% trimmed mean.
 For the situation
at hand, consider again $g=h=0$ and $Y=.5+ \epsilon$, only now method W is applied using means rather than 20\% trimmed means. Now power is estimated to .43, slightly better than
using a 20\% trimmed for which power was estimated to be .38. Using instead method G, power was estimated to be .56. So again, method G offers more power than method W and 
power is a bit higher compared to using a 20\% trimmed mean, which was .51.
For $(g, h)=(0,.2)$,  now the power of method W  was estimated to .25 when using a mean compared to .48 when using a 20\% trimmed mean. 
More relevant to the present paper is that if method G is used with a mean, power is estimated to be .30, which is substantially smaller than the estimate of .51 when using a 20\% trimmed mean.

\section{Illustrations}

 There is the issue of whether method G can reject when method W does not when dealing with data from an actual study. 
 There is also the issue of whether comparing quartiles  makes
a practical difference. We report results relevant to these issues using data from the Well Elderly 2 study.

 A general goal in the Well Elderly 2 study was to assess the efficacy of an intervention strategy  aimed
at improving the physical and emotional health of older adults.   
(The data are available at http://www.icpsr.umich.edu/icpsrweb/landing.jsp.)
A portion of the study
 was aimed at understanding the impact of intervention on depressive symptoms as measured by the
  Center for Epidemiologic 
Studies Depressive Scale (CES-D).  The CES-D (Radloff, 1977) is sensitive to change in depressive 
status over time and has been successfully used to assess ethnically diverse older people (Lewinsohn et al., 1988; Foley et al., 2002). Higher scores indicate a higher level
of depressive symptoms.
Another dependent variable was the RAND 36-item Health Survey (SF-36), a measure of self-perceived physical health and mental well-being 
(Hays, 1993; McHorney et al., 1993). Higher scores reflect greater health and well-being.

 Before intervention and  six months  following intervention, 
 saliva samples were taken at four times over the course of a single day:  on rising, 30 min after rising, but before taking anything by mouth, before lunch, and before dinner. 
 Then samples were assayed for cortisol.
 Extant studies (e.g.,
 Clow et al., 2004; 
 Chida \& Steptoe, 2009) indicate that measures of
  stress are associated with the
 cortisol awakening response (CAR), which is defined as the change in cortisol concentration
that occurs during the first hour after waking from sleep. (CAR is taken to be the cortisol level upon awakening  minus the level of cortisol after the participants were awake for about an hour.)
Here, the goal is to compare males and females after intervention based on CES-D and SF-36 measures
using the CAR as a covariate. 
 
 To illustrate that in practice the global test can reject when method  W does not, and that comparing lower or upper quantiles can make a 
practical difference, consider the goal of comparing males and females  based on CES-D measures using CAR as a covariate. 
No differences are detected based on a 20\% trimmed mean or median when using method W as well as the global test proposed here.
This remains the case when comparing .25 quantiles using a bootstrap version of method W. 
But when using method G to compare the groups based on the .25  quantile, a significant difference is found . 
That is,  there was no significant difference between males and females based on a measure of
location intended to reflect the typical response.  But the results indicate that there is a sense in which males tend to have even lower CES-D scores than females.

For the SF-36, testing (\ref{null_glob}) based on the median, a significant difference is found at the .05 level. (There were 75 males and 171 females after eliminating missing values.) 
Figure 1 shows a plot of the regression lines where
the solid lines is the regression line for males. For the males there were 6 outliers among the CAR values and for the females there were 8 outliers (based on a boxplot), which 
were eliminated from the analysis and are not shown in Figure 1. (Eliminating outliers among the independent variable is allowed. It is eliminating outliers 
among the dependent variable that can cause technical problems.)
For the situation in Figure 1, a bootstrap version of method  W  indicates  significant differences when CAR is negative (cortisol increases shortly after awakening).
In practical terms, the results indicated that the typical males perceived health and well being scores are higher  among individuals whose cortisol levels increase after awakening. When cortisol decreases, no
significant difference between males and females is found. Moreover, there appears to be little or no association between the CAR and SF-36 among women. For men, again there is
no significant association when cortisol increases.  But when cortisol decreases, a negative association is found. (The slope differs significantly from zero, $p=.03$, when fitting a straight line regression via a 
generalization of the Theil--Sen estimator that is designed to handle tied values.)

\begin{figure}
\resizebox{\textwidth}{!}
{\includegraphics*[angle=0]{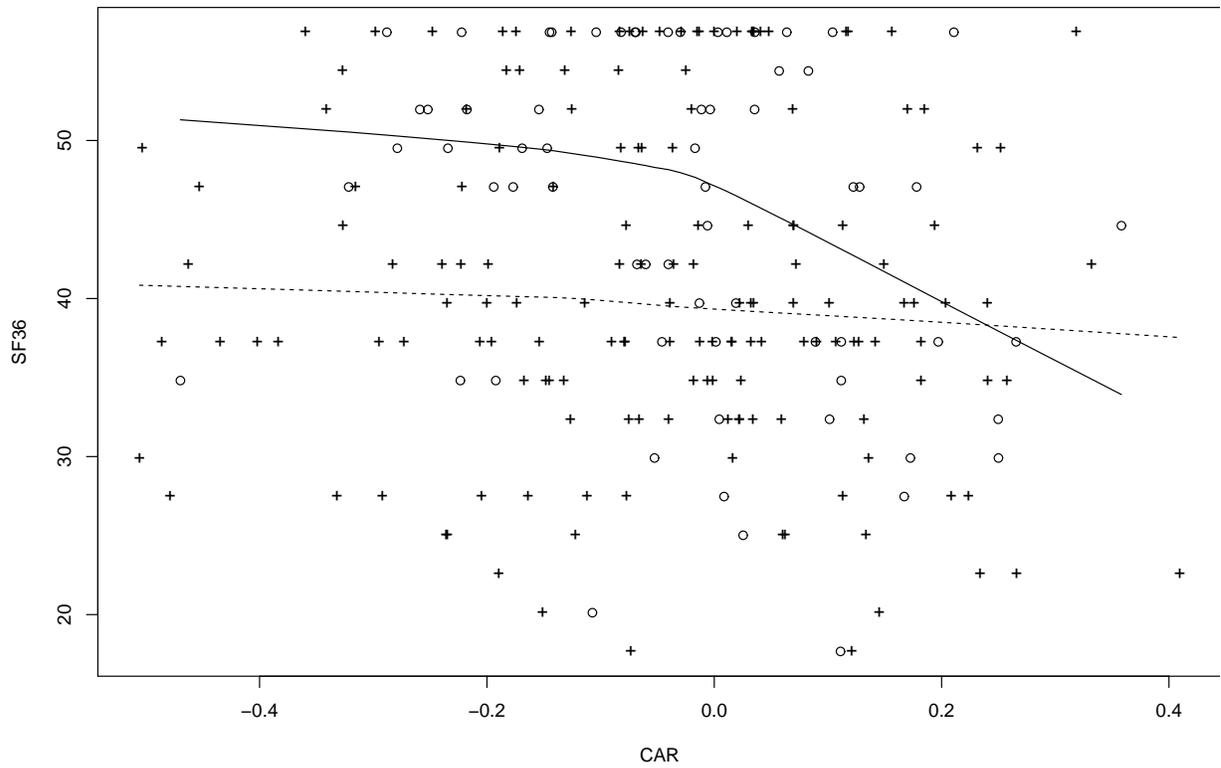}} 
\caption{Regression lines for predicting perceived health and well-being. The independent variable is the cortisol awakening response.  The solid line is the  .5 quantile regression line for males.}
\end{figure}


Note that in Figure 1, there appears to be curvature for the males. A test of the hypothesis that the regression line is straight was performed using the R function qrchk  in Wilcox (2012b, p. 544). 
 If again  the six outliers among the independent variable are eliminated, the hypothesis  of a straight line is rejected at the .05 level ($p=.046$). 
If the outliers are retained, now $p=.005$. 
So the results suggest that as CAR increases, there is little change in the typical SF-36 value when CAR is negative. But for CAR positive, the typical SF-36 value for males decreases.

To add perspective, Figure 2 shows the least squares regression lines for the same data used in  Figure 1.
If the classic ANCOVA method is applied, the slopes do not differ significantly at the .05 level ($p=.16$) and intercepts do differ significantly ($p=.008$). 
 But Figure 1 suggests that there is a distinct bend approximately where CAR is equal to $-.1$. Indeed, the least squares estimates of slope for males, based on the CAR values greater $-.1$, 
 differs 
 significantly from the slope for females  using a method that allows heteroscedasticity, $p=.011$. (Heteroscedasticity was addressed be estimating the standard errors via the HC4 estimator.
 See for example Wilcox, 2012a, p. 242. Again CAR values flagged as outliers by a boxplot  were removed.)  Using instead the Theil--Sen estimator, again the slopes are significantly different, $p=.047$.


\begin{figure}
\resizebox{\textwidth}{!}
{\includegraphics*[angle=0]{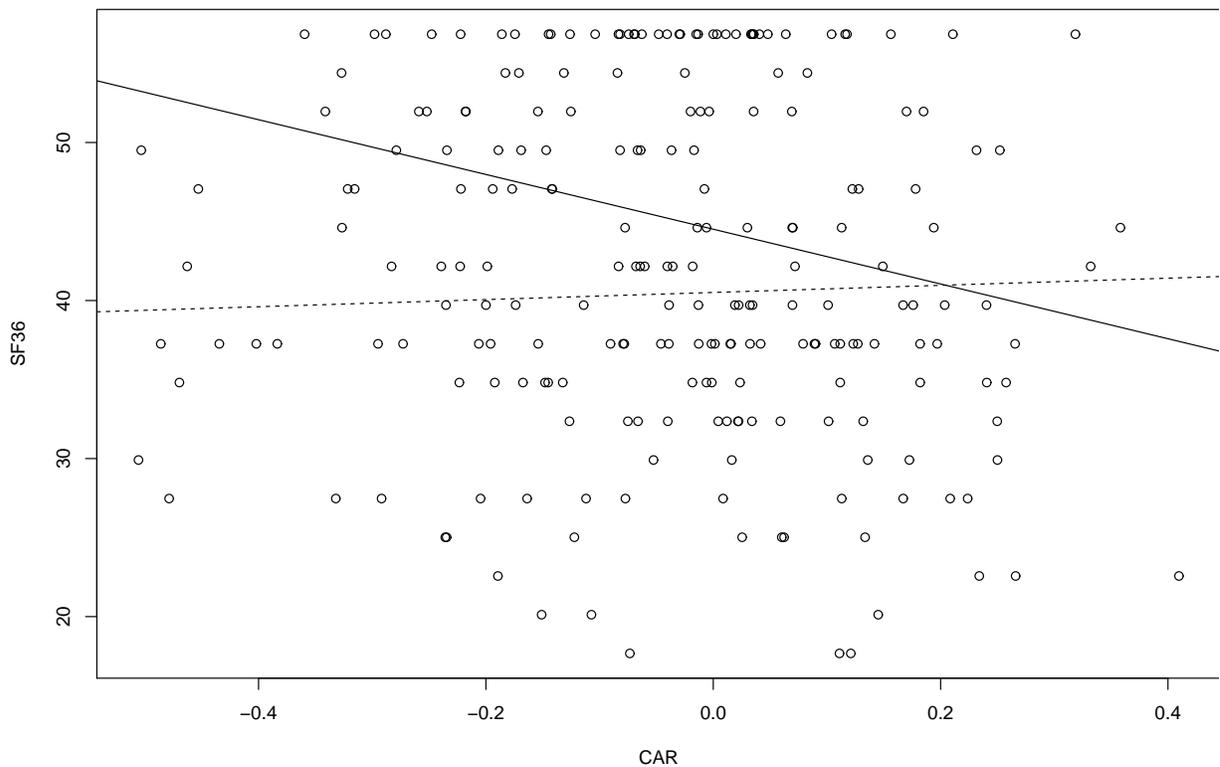}} 
\caption{The least squares regression lines for predicting perceived health and well-being using the same data shown in Figure 1.  Again, the solid line is the  regression line for males.}
\end{figure}

\section{Concluding Remarks}

In summary, all indications are that method G avoids Type I errors well above the nominal level. The highest estimated Type I error probability was .056 when testing at the .05 level.
The only known concern is that when dealing with a very heavy-tailed distribution, the Type I error probability might be less than .025 with  sample sizes of 30. Increasing the sample
sizes to 50, this problem was avoided among the situations considered. 

It is unclear under what circumstances some asymptotic result might be used to determine an appropriate critical value. The answer depends on the sample sizes, the span used by
the running interval smoother  ($\ell_1$ and $\ell_2$) and the
number of covariate values used. But this would seem to be a minor inconvenience in most situations because  a critical value
can be determined fairly quickly using the method described in
 the paper. Even with sample sizes of 300, execution time was only 39.5 seconds on a MacBook Pro.

It is not being suggested that method G dominates all  approaches relevant to ANCOVA. It seems fairly evident that no single method dominates, one reason being that different
methods are sensitive to different features of the data. Rather, method G provides an approach to ANCOVA that might have practical value in various situations, as was illustrated 
using the Well Elderly data. Here, for example, by dealing with curvature in a flexible manner, coupled with a robust measure of location, the results indicated that when CAR is
negative, typical SF-36 scores for males tend to be higher than scores for females. The extent they differ appears to have little to do with the value of CAR.
But  for CAR greater than zero, this is no longer the case.  The differences between males and females tend to decrease as
CAR increases. Both classic ANCOVA and robust methods indicate that males tend to have higher SF-36 scores. But the robust methods provide
a more detailed picture regarding when this is  this case. 
Method G is just one tool that helps provide a more detailed understanding of data beyond the non-robust and less flexible approach based on classic ANCOVA methods.
Put in broader terms, is there a single number or a single method that tells us everything we would like to know about how groups compare? We would suggest that the answer is no.
Method G is aimed at dealing with this issue.


Finally, R software is available for applying method G.  The function ancGLOB performs the calculations and is stored on the first author's web page.  For faster execution time, C++ 
subroutines have been written that compute the critical p-value. To take advantage of these subroutines, first  install the R package devtools with the R command 
install.packages(``devtools").
Then the C++  subroutines can be installed with the following commands:
\begin{verbatim}
library("devtools")
install_github( "WRScpp", "mrxiaohe")  
\end{verbatim}
Finally, when using the R function ancGLOB, set the argument cpp=TRUE.


ACKNOWLEGMENT 

We thank Xiao He for supplying the C++ code used in this study.

\begin{center}
REFERENCES
\end{center}

Bradley, J. V. (1978) Robustness? {\em British Journal of Mathematical and}  {\em Statistical Psychology, 31}, 144--152.

Brunner, E., Domhof, S \& Langer, F. (2002). {\em Nonparametric Analysis of} {\em Longitudinal Data in Factorial Experiments}. New York: Wiley.

Brunner, E. \& Munzel, U. (2000). The nonparametric Behrens-Fisher problem:
 asymptotic theory and small-sample approximation. {\em Biometrical Journal}, {\em 42}, 17--25.

Chida, Y. \& Steptoe, A. (2009). Cortisol awakening response and psychosocial factors: A  systematic review and meta-analysis.
 {\em Biological Psychology, 80}, 265--278.

Clark, F., Jackson, J., Carlson, M., et al. (2011). Effectiveness of a lifestyle intervention in promoting
the well-being of independently living older people:
results of the Well Elderly 2 Randomise Controlled Trial.  {\em Journal of Epidemiology and Community Health, 66},
 782--790. doi:10.1136/jech.2009.099754

Cliff, N. (1996). {\em Ordinal Methods for Behavioral Data Analysis}.  Mahwah, NJ: Erlbaum.

Clow, A., Thorn, L., Evans, P. \&  Hucklebridge, F. (2004). The awakening cortisol response: Methodological issues and significance.
{\em Stress, 7}, 29--37.

Dielman, T., Lowry, C. \& Pfaffenberger, R. (1994). A comparison of quantile estimators. {\em Communications in Statistics--Simulation and}
{\em Computation, 23}, 355-371.

Doksum, K. A., \& Sievers, G. L. (1976). Plotting with confidence: graphical comparisons of two populations. {\em Biometrika, 63}, 421--434.

Dunnett, C. W. (1980). Pairwise multiple comparisons in the unequal variance case. {\em Journal of the American Statistical Association, 75}, 796--800.

Eakman, A. M., Carlson, M. E. \& Clark, F. A. (2010). The meaningful activity participation assessment: a measure
 of engagement in personally valued activities {\em International  Journal of Aging Human  Development, 70}, 299--317.

Efromovich, S. (1999). {\em Nonparametric Curve Estimation: Methods, Theory and 
 Applications}. New York: Springer-Verlag.

Efron, B. \& Tibshirani, R. J. (1993). {\em An Introduction to the Bootstrap}. New York: Chapman \& Hall.

Eubank, R. L. (1999). {\em Nonparametric Regression and Spline Smoothing}. New York: Marcel Dekker.


Foley K., Reed P., Mutran E., et al. (2002). Measurement adequacy of the CES-D  among a sample of older African Americans. {\em Psychiatric Research, 109}, 61--9.

Fox, J. (2001). {\em Multiple and Generalized Nonparametric Regression}.  Thousands Oaks, CA: Sage

Fung, K. Y. (1980). Small sample behaviour of some nonparametric multi-sample  location tests in the presence of dispersion differences. {\em Statistica} {\em Neerlandica, 34}, 189--196.

Gy\"{o}rfi, L., Kohler, M., Krzyzk, A. \&  Walk, H.  (2002). {\em A Distribution-Free Theory of Nonparametric Regression}. New York: Springer Verlag.

Hampel, F. R., Ronchetti, E. M., Rousseeuw, P. J. \& Stahel, W. A. (1986).
 {\em Robust Statistics}. New York: Wiley.

H\"{a}rdle, W., 1990. Applied Nonparametric Regression. Econometric
 Society Monographs No. 19, Cambridge, UK: Cambridge University Press.

Harrell, F. E. \& Davis, C. E. (1982). A new distribution-free quantile estimator. {\em Biometrika, 69}, 635--640.

Hays, R. D., Sherbourne, C .D. \& Mazel, R. M. (1993). The Rand 36-item health survey 1.0. {\em Health Economics, 2}, 217--227.

Heritier, S., Cantoni, E, Copt, S. \& Victoria-Feser, M.-P. (2009). {\em Robust Methods in Biostatistics}. New York: Wiley.

Hettmansperger, T. P. (1984). {\em Statistical Inference Based on Ranks}.  New York: Wiley.

Hettmansperger, T. P. \& McKean, J. W. (1998). {\em Robust Nonparametric} {\em Statistical Methods}. London: Arnold.

Hoaglin, D. C. (1985). Summarizing shape numerically: The g-and-h distribution. In D. Hoaglin, F. Mosteller \& J. Tukey (Eds.) {\em Exploring Data Tables Trends and Shapes}. New York: Wiley, pp. 461--515.

Hochberg, Y. (1988). A sharper Bonferroni procedure  for multiple tests of significance. {\em Biometrika, 75}, 800--802.

Huber, P. J. \& Ronchetti, E. (2009). {\em Robust Statistics}, 2nd Ed. New York: Wiley.

Jackson, J., Mandel, D., Blanchard, J., Carlson, M., Cherry, B., Azen, S., Chou, C.-P.,  Jordan-Marsh, M., Forman, T., White, B., Granger, D., Knight, B., \& Clark, F. (2009). Confronting challenges in intervention research with ethnically diverse older adults: the USC Well Elderly II trial. {\em Clinical Trials, 6}  90--101.

Kendall, M. G., \& Stuart, A. (1973). {\em The Advanced Theory of Statistics}, Vol. 2. New York: Hafner.

Lawson, A. (1983). Rank Analysis of Covariance: Alternative Approaches. {\em Statistician, 32},  331--337.

Lewinsohn, P.M., Hoberman, H. M., Rosenbaum M. (1988). A prospective study of risk factors  for unipolar depression. {\em Journal of  Abnormal Psychology, 97}, 251--64.

Liu, R. G. \& Singh, K. (1997). Notions of limiting P values based on data
 depth and bootstrap. {\em Journal of the American Statistical Association, 92},
 266--277.

Lombard, F. (2005). Nonparametric confidence bands for a quantile comparison
 function.  Technometrics, 47, 364--369.

 Maronna, R. A., Martin, D. R. \& Yohai, V. J. (2006). {\em Robust Statistics:}
 {\em Theory and Methods}. New York: Wiley.
 
 McHorney, C. A., Ware, J. E. \& Raozek, A. E. (1993). The MOS 36-item Short-Form Health Survey (SF-36): II. Psychometric and clinical tests of validity in measuring physical
 and mental health constructs. {\em Medical Care, 31}, 247--263.

Neuh\"{a}user, M., L\"{o}sch, C., \& J\"{o}ckel, K-H (2007). The Chen-Luo  test in case of heteroscedasticity. {\em Computational Statistics \&} {\em Data Analysis, 51}, 5055--5060.

Parrish, R. S. (1990). Comparison of quantile estimators in normal sampling.
 {\em Biometrics, 46}, 247--257.

Radloff, L., 1977. The CES-D scale: a self report depression scale for research in the general population. {\em Applied Psychological Measurement 1}, 385--401.

Rom, D. M. (1990). A sequentially rejective test procedure based on a modified  Bonferroni inequality. {\em Biometrika, 77}, 663--666.

Rosenberger, J. L. \& Gasko, M. (1983). Comparing location estimators: Trimmed
 means, medians, and trimean. In D. Hoaglin, F. Mosteller and J. Tukey (Eds.) {\em Understanding Robust and exploratory data analysis.} (pp. 297--336). New York: Wiley.

Sfakianakis, M. E. \& Verginis, D. G. (2006). A New Family of Nonparametric Quantile Estimators.
{\em Communications in Statistics--Simulation and Computation, 37}, 337--345.

Sheather, S. J. \& Marron, J. S. (1990). Kernel quantile estimators. {\em Journal}
 {\em of the American Statistical Association, 85}, 410--416.

Staudte, R. G. \& Sheather, S. J. (1990).
{\em Robust Estimation and Testing}.  New York: Wiley.


 Wilcox, R. R. (1997). ANCOVA based on comparing a robust measure of
 location at empirically determined design points. {\em British}
 {\em Journal of Mathematical and Statistical Psychology},
      {\em 50}, 
 93--103.

Wilcox, R. R. (2009).  Robust ANCOVA using a smoother with bootstrap  bagging. {\em British Journal of Mathematical and Statistical              
 Psychology}  {\em 62}, 427--437.

Wilcox, R. R. (2012a). {\em Introduction to Robust Estimation and
Hypothesis Testing}, 3rd Ed. San Diego, CA: Academic Press.

Wilcox, R. R. (2012b).  Modern Statistics for the Social and Behavioral Sciences:  A Practical Introduction. New York: Chapman \& Hall/CRC press

Yuen, K. K. (1974). The two sample trimmed t for unequal population variances.
 {\em Biometrika, 61}, 165--170.

\end{document}